\documentclass[aps,pra,twocolumn,groupedaddress,nofootinbib]{revtex4-1}
\usepackage{graphicx}
\usepackage{subfigure}
\usepackage{mathtools}
\usepackage{empheq}
\usepackage{multirow}
\usepackage{color}
\usepackage[table,x11names]{xcolor}

\begin{document}


\title{Optical levitation using broadband light}

\author{A. T. M. Anishur Rahman}
\email{a.rahman@ucl.ac.uk}
\author{P. F. Barker}
\email{p.barker@ucl.ac.uk}
\affiliation{Department of Physics and Astronomy\\
	University College London\\
	Gower Street, WC1E 6BT
	London, UK} 


\date{\today}

\begin{abstract}
The ability to create dynamic, tailored optical potentials has become important across fields ranging from biology to quantum science. We demonstrate a method for the creation of arbitrary optical tweezer potentials using the broadband spectral profile of a superluminescent diode combined with the chromatic aberration of a lens. A tunable filter, typically used for ultra-fast laser pulse shaping, allows us to manipulate the broad spectral profile and therefore the optical tweezer potentials formed by focusing of this light. We characterize these potentials by measuring the Brownian motion of levitated nanoparticles in vacuum and, also demonstrate interferometric detection and feedback cooling of the particle’s motion. This simple and cost-effective technique will enable wide application and allow rapid modulation of the optical potential landscape in excess of MHz frequencies.    
\end{abstract}


\maketitle

\section{Introduction}


\indent Tailored optical tweezer potentials find application in fields ranging from biology and chemistry, to rheology and atom optics\cite{Folling2007, WoerdemannMike2013Aotb,Gauthier2016,Stuart_2018, CURTIS2002,Schonbrun2005,Boyer2006PRA,Preece_2011,ZhangH2008}. More recently, levitation of nanoparticles in vacuum using optical tweezers in combination with the control of the center-of-mass temperature has been used to explore quantum mechanics in a new high mass regime \cite{TebbenjohannsPRL2020,Delic2019}. Here, the creation of well-controlled and rapidly modulated non-linear potentials is seen as a promising route to explore their quantum, non-classical motion.

\indent Exquisite control over the phase and/or the amplitude of the light field, via spatial light modulators or digital mirror devices, has allowed the creation of complex optical potentials that can be changed over sub millisecond times scales \cite{Boyer2006PRA,WoerdemannMike2013Aotb,Gauthier2016,Stuart_2018}. Other methods utilise rapid scanning of a single field to create time averaged tailored potentials \cite{BerutNature2012}. These optical potentials are typically created by using a strong monochromatic laser which can be tightly focused due to its narrow linewidth and high spatial coherence.  Generally, however, broadband incoherent light sources are not considered to be useful for optical tweezers in vacuum because they are typically of low intensity and are subject to chromatic aberration preventing the tight focusing required for the creation of deep optical potentials.  However, traps have been constructed from broadband light sources using supercontinuum or femotosecond lasers. They have been shown to have extended focal regions due to the dispersion in the lens coupled with the large spectral spread of these sources. The broad spectral bandwidth can also be used for guiding particles and for spectroscopy of the trapped object \cite{Agate2004,Li2005OL,Fischer:06}. A holographic white light  trapping  system  has demonstrated both trapping and rotation of particles using a supercontinuum vortex beam \cite{Morris:08}. Such sources have also been used to transfer orbital angular momentum to the trapped particles \cite{Wright:08}. In addition, the high peak intensities of these sources, have been shown to be useful for inducing non-linear optical properties within the trapped particle \cite{Gong:18}.

\begin{figure}
	\subfigure{\includegraphics[width=8.25cm]{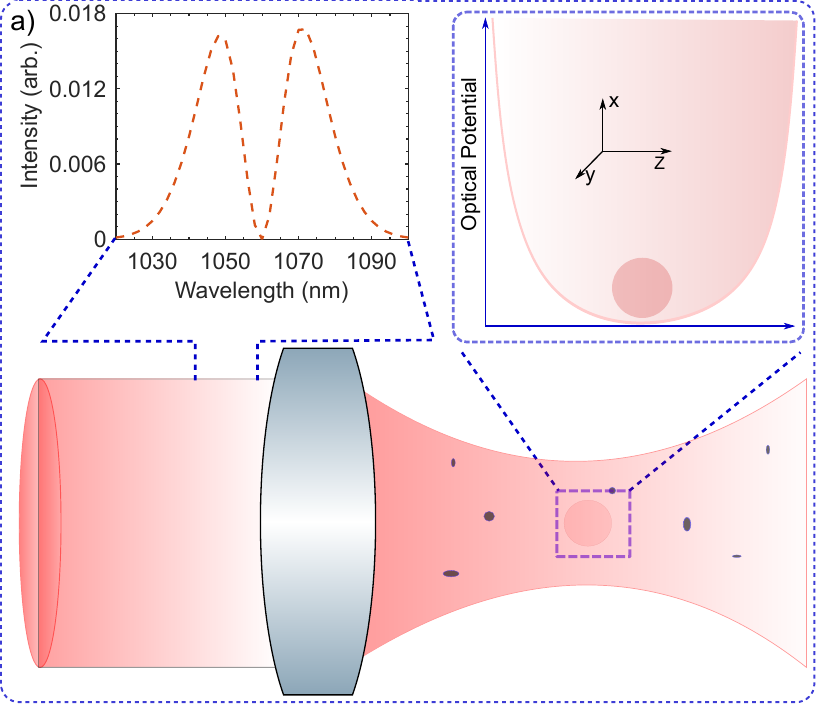}}
	\subfigure{\includegraphics[width=8.25cm]{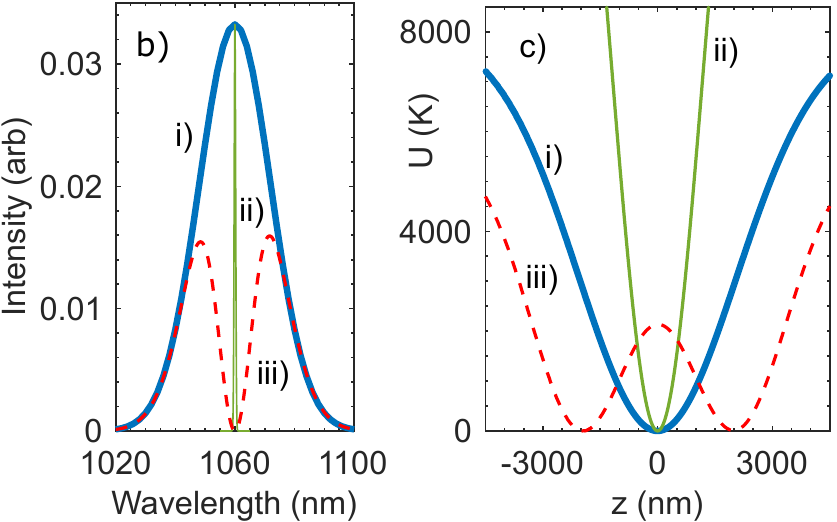}}
	\subfigure{\includegraphics[width=8.25cm]{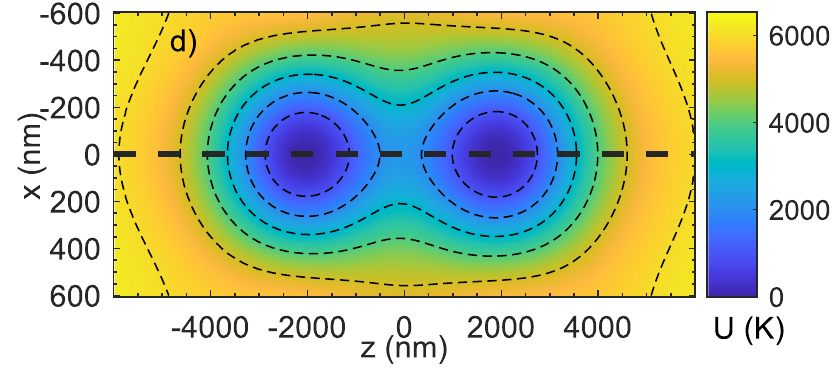}}
	\caption{\label{fig2c} a) A simplified experimental schematic where an aspheric lens made of lanthanum flint glass (D-ZLaF52LA) is used to form an optical tweezer trap for levitation. The left inset shows the spectral profile of the input light. The right inset is a sketch of the potential profile that can be generated using the chromatic aberration of the trapping lens and the filtered spectral profile of the trapping light (see text for details). b) Profiles produced by - i) a broadband light source with a linewidth of $\approx 28~$nm, ii) a laser source of linewidth $0.10~$nm and iii) a filtered profile obtained from the broadband source i). c) The potential wells associated with the spectral profiles of part b). The minima of all potential profiles have been set to zero for the purpose of comparison. d) The potential landscape in the x-z plane for the intensity profile $\romannumeral 3$ of part b). The potential profile $\romannumeral 3$ of part c) is along the thick dashed black line. Contour lines are equipotentials. In calculating the potentials we use a silica nanosphere of $R = 50~$nm, a trapping power of $300~$mW at the entrance of the lens, a lens diameter of $5~$mm and a beam diameter of $8~$mm at the entrance of the lens. The lens has a focal shift of $150~$nm per $1~$nm change in the wavelength.}
\end{figure}
\indent Superluminescent diodes (SLD) are relatively inexpensive sources of intense, broadband light that are produced by amplified spontaneous emission \cite{CahillAPL2019}. These devices have large linewidths that range from a few nanometers up to $100~$nm and their low coherence finds application in a large variety of applications including optical coherence tomography \cite{KoOpticExp2004} and fiber optic gyroscopes \cite{Bohm1981}. Although SLDs have poor temporal coherence, they have high transverse spatial coherence when coupled into a single mode fiber, allowing light to be focused to the small spot sizes required for optical trapping. Finally, as powers that exceed $100$s of mW can be coupled into the fiber, there is sufficient power to form deep optical traps. 

\indent In this article, we demonstrate that the spectrally broad light of a superluminescent diode can be used to form deep and stable optical trapping potentials which are capable of levitating particles in vacuum. Importantly, we show that by filtering this light source using a tunable spectral filter constructed of a simple grating pair, that the inherent chromatic aberration of a typical lens allows us to produce tunable linear and non-linear optical potentials. In addition, we show that parametric feedback cooling can be undertaken to control the center-of-mass temperature of a levitated particle and is as effective as that carried out by a conventional laser trap.

\section{Optical potential from a broadband light source}

The optical tweezers potential for subwavelength nanospheres is dominated by the dipole force which is determined by the intensity profile of the focused light beam $I(r)$, the dielectric constant $\epsilon$, and volume $V$ of the particle. The optical dipole potential \cite{SkeltonDholakia2016} in vacuum is given by 
$U(r)=-\frac{3V}{2c}\Re{\Big(\frac{\epsilon-1}{\epsilon+2}\Big)}I(r)$, where $c$ is the speed of light in vacuum. The Gaussian spatial profile created by a focused laser beam produces a Gaussian potential well. However, when the energy of the particle is much less than the well depth it is very well approximated by a quadratic potential creating a simple harmonic oscillator in all three dimensions. The optical potential produced by a broadband source such as a superluminescent diode can be significantly different to that produced by a single mode laser since chromatic aberration due to dispersion in the focusing lens leads to different focal lengths and focused spot sizes for each wavelength component. This is schematically illustrated in Fig. \ref{fig2c}a in which a high numerical aperture lens with chromatic aberration focuses a spectrally filtered broadband source, such that at the focus, a non-Gaussian or non-linear potential is created. This is further highlighted in Figure \ref{fig2c}b~$\&$~c which shows three spectral profiles and their corresponding potentials calculated using the dispersion of a commercially available aspheric lens with a numerical aperture of N.A.=0.77 (D-ZLaF52LA, Edmund optics, dispersion of $dn/d\lambda = $-0.024127 $\mu m^{-1}$ at a wavelength of 1.06 $\mu m$ ). The intensity profile in the focus is calculated using the Richards-Wolf formalism \cite{NovotnyHecht2012} and the contribution of all wavelength components is expressed as an incoherent sum as the interference terms between all of the different fields averages out to zero. Here the intensity is given by $I(r)=\frac{\epsilon_0 c}{2}\sum_{\lambda_{min}}^{\lambda_{max}}w_i\Bigl(E_x(r,\lambda_i)^2+E_y(r,\lambda_i)^2+E_z(r,\lambda_i)^2\Bigr)$, where $w_i$ is the spectral weight, and $E_x(r,\lambda_i)$, $E_y(r,\lambda_i)$ and $E_z(r,\lambda_i)$ are the position and wavelength dependent optical field at position $r$ around the focus of the trapping lens along the three major axes. The focal length of the lens for each wavelength along the direction of light propagation ($z-$axis) is determined by the dispersion of the lens material, with a nominal effective focal length of 3.1 mm. Figure \ref{fig2c}c shows the derived potentials in units of Kelvin ($U/k_B$), corresponding to the three experimentally feasible intensity profiles shown in Fig. \ref{fig2c}b. Here, $k_B$ is the Boltzmann constant. Our simulation has a spectral resolution of $1~$nm and a spatial resolution of $10~$nm. Due to the dispersion of our lens and diffraction, simulations with spectral resolution of up to $5~$nm and spatial resolutions of less $50~$nm do not significantly change the profiles shown here. The full-width-half-maximum linewidth of the broadband source (profile \romannumeral 1, Fig. \ref{fig2c}b) is $28~$nm. This is the minimum linewidth required, in combination with spectral filtering, that will create a double well potential using the chromatic aberration of our lens. The minima of all potential profiles in Fig. \ref{fig2c}c have been set to zero for the purpose of comparison. In the simulation we have used a $R=50~$nm silica nanoparticle, as these are later used in our experiment. We use a trapping power of $300~$mW at the entrance of the trapping lens corresponding to approximately $150~$mW in the trapping region due to apodization. Additionally, a focal shift of $150~$nm per $1~$nm change in the trapping wavelength associated with the dispersion of the lens material has been used. As expected, the optical potential produced by the broadband source with a linewidth of $28~$nm (profile $\romannumeral 1)$, is much shallower than that produced by an ideal single mode laser source (profile $\romannumeral 2$) but is still deep enough for trapping particles($> 10k_BT$ \cite{NovotnyHecht2012}). This lower well depth (profile $\romannumeral 1$) arises because each wavelength component focuses at a different location along the $z-$axis creating a spread in the focus. The intensity profile $\romannumeral 3$ in Fig. \ref{fig2c}b, generated from the spectral profile $\romannumeral 1$ using a notch filter, produces a double-well potential (profile $\romannumeral 3$, Fig. \ref{fig2c}c). The formation of this potential can be understood from spectral profile $\romannumeral 3$, in which there are effectively two peaks. On focusing this light using a high NA aspheric lens, its chromatic aberration produces two spatially separated focal spots which form two potential wells. Here, the well near $z\approx -2000~$nm ($z\approx 2000~$nm) is associated with the peak centered around $1045~$nm ($1070~$nm) (profile $\romannumeral 3$, Fig. \ref{fig2c}b). The depth of each of the potential wells is $\approx 2000~$K while the separation between the minima of the two wells is $4000~$nm. Note that in order to achieve the same well depths the longer wavelength spectral peak at $1070~$nm must have a higher relative intensity than that at $1048~$nm to compensate for the change in focused spot size with wavelength. The depth of each individual well, their separation and barrier height can be modified by changing the linewidth and the center wavelength of the notch filter (see below for more details). Of course, as the bandwidth is increased we must increase the SLD power to maintain the well depth for trapping. Figure \ref{fig2c}d shows a $2$D plot of the potential landscape in the $x-z~$plane corresponding to the intensity profile $\romannumeral 3$, Fig. \ref{fig2c}b. Two wells (blue areas) can be seen and the contour lines represent equipotentials. In this simulation, the polarization of the light was along the $y-$axis. The potential along the thick dashed black line in Fig. \ref{fig2c}d is equivalent to profile \romannumeral 3, Fig. \ref{fig2c}c. The depth of the potential well along the $x-$axis at $ z=\pm2000~$nm is $\approx 2000~$K and symmetric about the $z-$axis.

\begin{figure}[!t]
	\includegraphics[width=8.500cm]{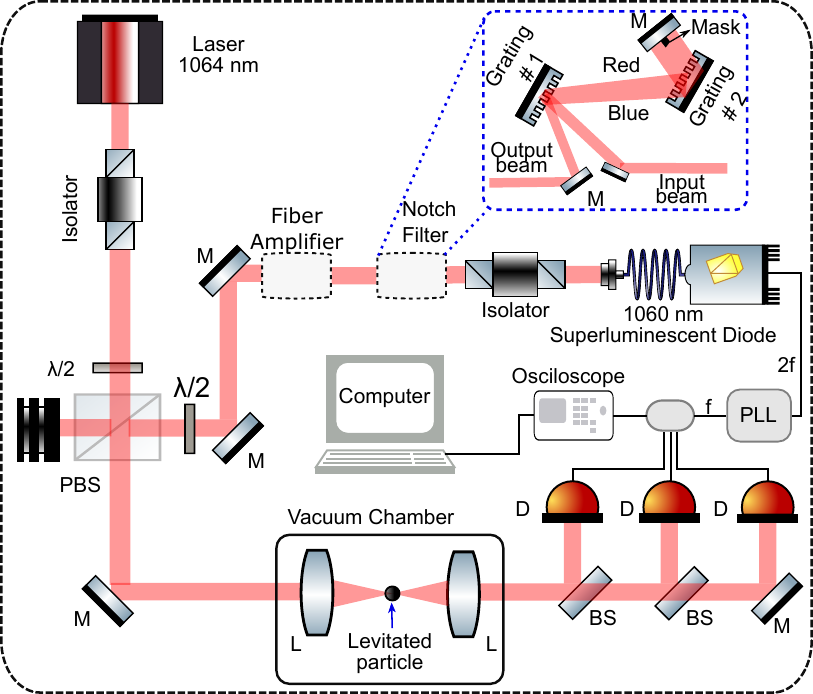}
	\caption{\label{fig1} Optical layout used for levitation. The labelled components are : $\lambda/2$ - half waveplate, M - mirror, BS - beam splitter, PBS - polarizing beam splitter, D - balanced photodiode and L - lens. Dotted lines around components denote parts that are used for some of the experiments. Inset shows the home-built notch filter which consists of a beam block, a mirror and two identical gratings mounted parallel to each other. The linewidth and the center wavelength of the notch filter can be tuned by changing the width and the position of the block. For parametric feedback using the SLD, signals from the balanced photodiodes are fed to phase-locked loops (PLL). The output of the PLL, with suitable attenuation, is used as the input to the SLD current controller. See main text for more details.}
\end{figure}

\begin{figure}[t]
	\includegraphics[width=8.50cm]{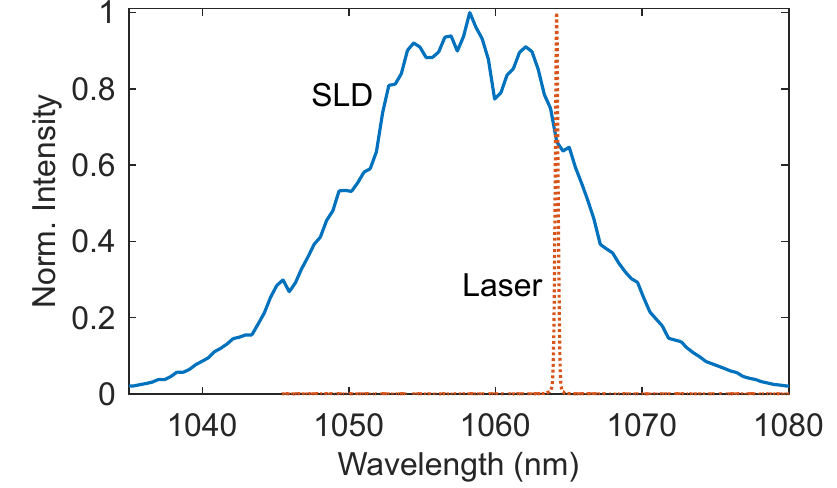}
	\caption{\label{fig3} The spectral profile of an unfiltered and unamplified superluminescent diode and a single mode Nd:YAG laser. Note that the laser has a significantly narrower linewidth than shown here. This is due to the finite resolution ($0.10~$nm) of our spectrometer.}
\end{figure}

\section{Experimental setup}

In our experiment an optical tweezers is formed using a $0.77$ numerical aperture (NA) aspheric lens made of dense lanthanum flint glass (D-ZLaF52LA, part no. 83-674, Edmund Optics) with a focal length of $3.1~$mm. The lens is housed inside a vacuum chamber as shown in Fig. \ref{fig1} and levitation of silica nanospheres is carried using either a CW super luminescent diode or a CW Nd:YAG laser. The SLD wavelength is centered around $1060$ nm and has a linewidth of $18~$nm and is shown in Figure \ref{fig3}. The SLD does have a stable ripple structure indicating that it is not completely modeless. Also shown in Fig. \ref{fig3} is the spectral profile of the $1064~$nm Nd:YAG single mode laser output whose linewidth is significantly lower ($\approx 10$ kHz) than shown. This is due to the finite resolution of $0.10~$nm of our Andor Shamrock 303 spectrometer. The beams from each source are combined on a polarizing beam splitter (PBS) and propagate co-linearly into the trapping lens. To compensate for losses via spectral filtering, we use a fiber amplifier to amplify the beam up to a maximum of $1.4~$W. When the notch filter (see Fig. \ref{fig1}) is used the power of the trapping beam remains fixed irrespective of the spectral profile of the input beam due to saturation of the amplifier gain.

The tunable notch filter, which is used to create the non-linear optical potential, is discussed in more detail below. For some spectral profiles we can trap particles without the amplifier. Once levitated, we detect the particle's oscillatory motion in each trap axis using three balanced photodiodes \cite{Gieseler2012,LiNatPhys2011}. For parametric feedback cooling, the signals from the photodiodes are fed to lock-in amplifiers where internal oscillators are phase locked to each of the three trap frequencies. The sum of the three oscillators is fed to the current controller of the superluminescent diode. This modulates the output of the SLD generating the signal for the parametric feedback cooling. Modulating the current directly means that we do not require an acousto/electro optic modulator \cite{Gieseler2012,Vovrosh2017}. For typical operation, the modulation index of the intensity fluctuation of the SLD was less than $1\%$.

The SLD spectral profile is modified by a notch filter (see the inset, top right corner, Fig. \ref{fig1}) consisting of a retro-reflecting mirror and two identical blazed diffraction gratings ($600~$grooves/mm, Thorlabs Inc.) mounted parallel to each other. This arrangement is typically used for compression of pulses in chirp pulse amplification schemes \cite{Agostinelli1979}. To operate this filter, the spectrally broad SLD beam, as shown in Fig. \ref{fig3}, is collimated and directed towards the first grating. This spectrally dispersed and diverging beam from the first grating is directed onto a second grating which is arranged to prevent further spectral dispersion creating a collimated beam. In this beam, the wavelength components are spatially dispersed in the horizontal plane (see Fig. \ref{fig1}). To modify the spectral profile we simply place a mask in the beam which blocks the appropriate spectral components. The width ($3-7~$mm) of the mask determines the spectral contents removed from the beam while its position with respect to the beam fixes the centre wavelength of the notch filter. The filtered beam is then retroflected by a mirror back through the grating pair where the spectral components are recombined into a collimated beam that can be used for trapping. The return beam is slightly displaced vertically with respect to the incoming beam and is picked off using a D-mirror as shown in figure \ref{fig1}. The return beam is then coupled into the optical amplifier to achieve the desired level of power for levitation.

\begin{figure}[!t]
	\subfigure{
		\includegraphics[width=8.5cm]{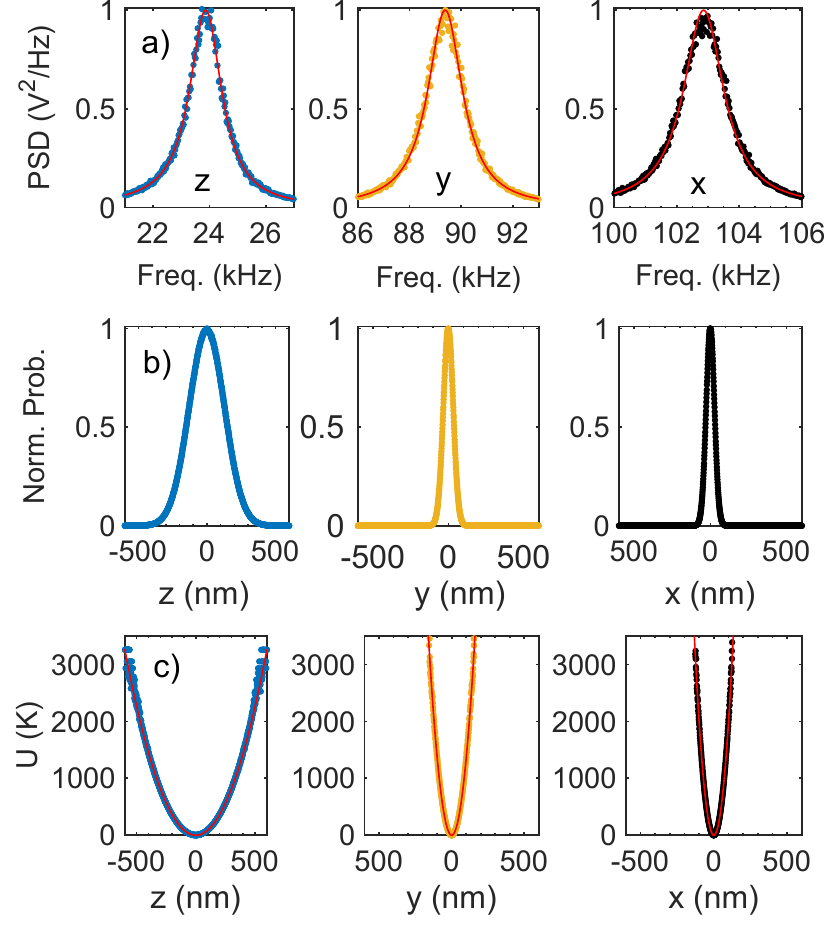} }
	\subfigure{
		\includegraphics[width=8.50cm]{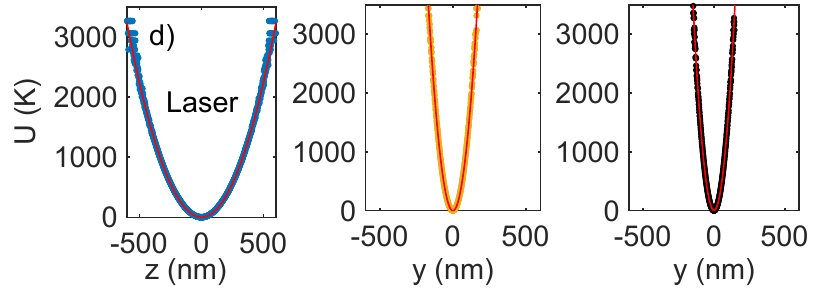} }
	\caption{\label{fig4} Levitation using the superluminescent diode. a) The power spectral density along the three principle axes: $z-$axis represents the direction of light propagation, $y-$axis is parallel to the direction of the electric field ($E$) polarization, and $x-$axis is perpendicular to $E$ field polarization. b) The normalized position histograms along the three axes obtained from the calibrated time traces. c) The potential profiles derived from the position histograms. d) Potential profiles along the three axes for the same particle used in parts a-c but under the laser levitation. The frequency along the $z-$axis was purposely made equal to that under the SLD levitation, part a. This experiment was performed at $\approx 2~$mBar. Red solid lines in parts c $\&$ d are quadratic functions. See main text for details.}
\end{figure}

\begin{figure}[!t]
	\subfigure{\includegraphics[width=8.50cm]{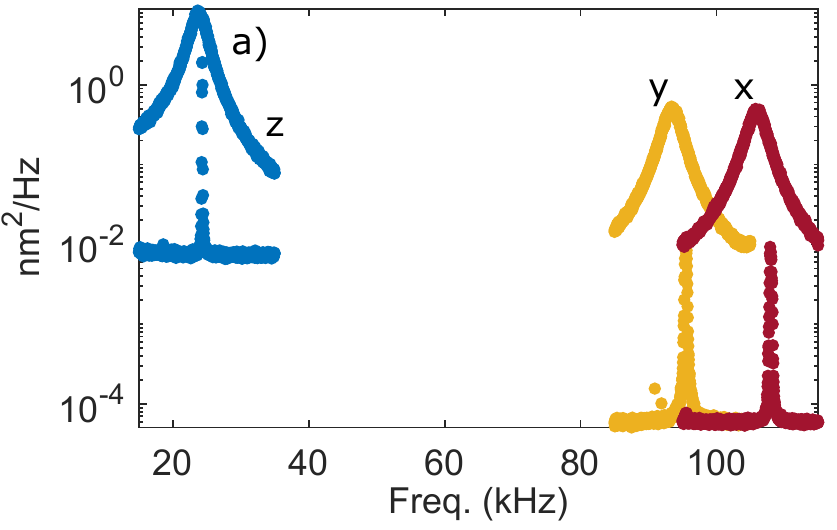}}
	\subfigure{\includegraphics[width=8.50cm]{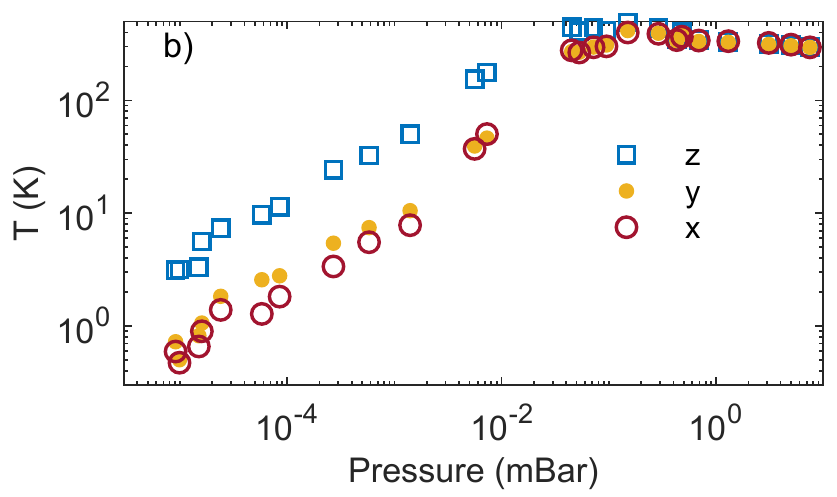}}
	\caption{\label{fig6} Parametric feedback cooling using the SLD when no spectral filtering is in use. a) Power spectral densities (PSD) along the three axes. The top graphs show PSDs at $5~$mBar when no feedback is applied while the bottom graphs are the PSDs under parametric feedback cooling at $\approx 9\times 10^{-6}~$mBar. b) The center-of-mass temperature along the three translational axes as a function of pressure under parametric feedback cooling.}
\end{figure}

\begin{figure*}[!t]
	\includegraphics[width=18.250cm]{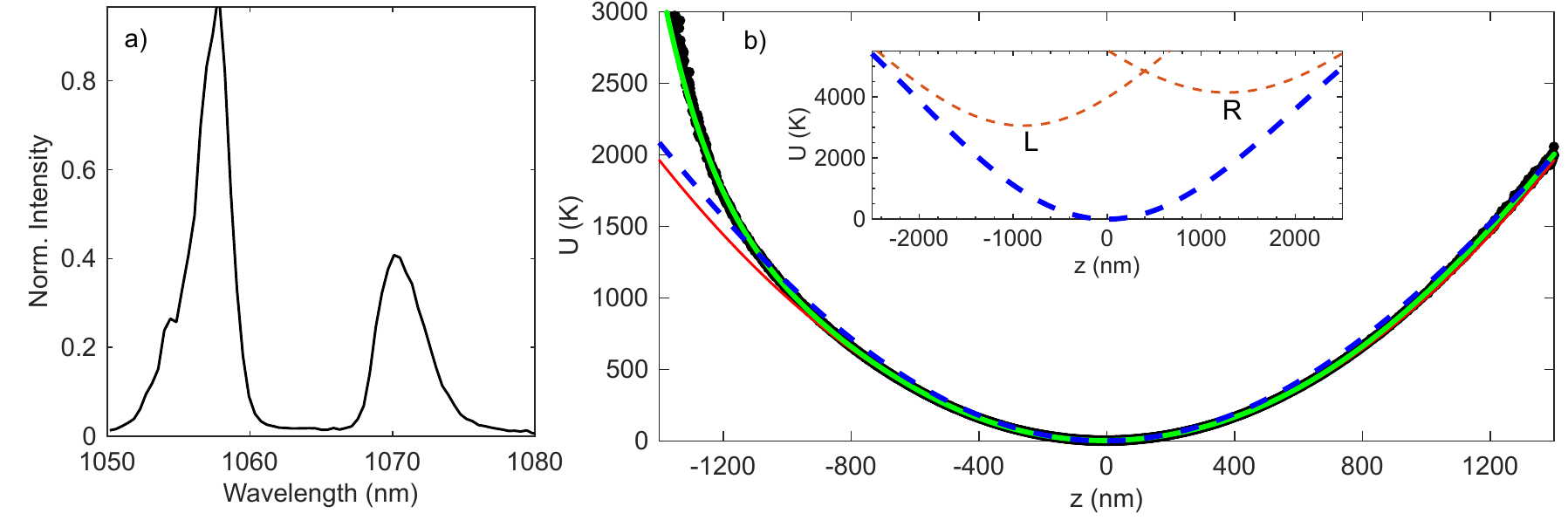}
	\caption{\label{fig5} Non-linear optical potential. a) SLD intensity profile after spectral filtering using a notch filter. b) Black dots are the experimental optical potential in unit of Kelvin. Experimental $U(z)$ has been obtained from the position histogram along $z-$axis. The red solid line represents a quadratic function. The blue dashed line is the simulation results obtained using the procedure outlined in Section 2 using the spectral profile of part a), particle radius $R=50~$nm, obtained from the linewidth measurement by transferring the particle to the laser beam \cite{RahmanSciRep2016}, and a trapping power of $150~$mW in the focus. The green line is a polynomial fit using $U(z)=a_0+a_1z+a_2z^2+...+a_9z^9$, where $a_i,$s are the fitting parameters and $z$ is in nanometer. The inset shows the left (L) and right (R) wells that correspond to the peaks at $1058~$nm and $1070~$nm (part a) and their sum (blue dashed line). See main text for more details.}
\end{figure*}

\section{Levitation using broadband light}

Nanoparticles were loaded into the trap by ultrasonic nebulization of silica nanoparticles dissolved in methanol at atmospheric pressure \cite{Gieseler2012, RicciNatComm2017}. The trapping power at the focus was approximately $150~$mW from the superluminescent diode without any amplification or filtering. Once trapped, the chamber pressure was rapidly reduced down to a pressure of $\approx 5~$mBar where the underdamped motion of the particles in the trap can be clearly resolved. At this pressure internal heating is not significant and the motional temperature of the particle can be well approximated by the room temperature value of $295~$K \cite{Gieseler2012,Vovrosh2017}. Figure \ref{fig4}a shows the averaged power spectral densities (PSD), derived from time traces of the balanced photodiodes, of a $100~$nm diameter silica nanoparticle (Corpuscular Inc.). The graphs are an average of $66$ PSDs taken over a duration of $10~$seconds with a sampling rate of $1~$MHz. The figures show that like laser based levitation \cite{Gieseler2012,Vovrosh2017,RahmanSciRep2016,RahmanNatPhot2017,RahmanRevScIns2018,GambhirPRA2015,LiNatPhys2011}, the oscillation frequency along the light propagation direction ($z$-axis) is the lowest while that in the direction ($x-$axis), orthogonal to the electric field polarization, is the highest. The difference in oscillation frequency between the $x-$axis and the $y-$axis (parallel to the direction of $E$ field) is due to the asymmetry of the focus that occurs due to non-paraxial focusing of linearly polarized light with a high NA lens \cite{NovotnyHecht2012}. To convert the voltage $V$ measured by the balanced photodiodes into position $q$, we assume the linear relationship, $V=C q$, which is a good approximation for the small displacements of the trapped particles from the trap center \cite{Neuman2004}. Here, $C$ is the calibration factor. The power spectral density of a record of the voltage from the detectors is given by $S_{VV}=C^2 S_{qq}$, where $S_{qq}=\frac{2 k_B T}{M}~\frac{\Gamma_0}{(\omega^2-\omega_0^2)^2+\Gamma_0^2\omega^2}$ is the PSD in position in any axis with trap frequency $\omega_0/2\pi$. Here $M$ is the mass of the particle which is determined from the density $\rho$ and particle radius $R$, $T$ is the center-of-mass (cm) temperature, taken here to be $295~$K, and $\Gamma_0$ is the linewidth of the particle. To retrieve $C$ from the fitting parameter $C^2\frac{2k_B T}{M}$, one needs to know mass $M=4\pi R^3 \rho/3$. For the density of the particle \cite{Jain2016} we take $2200~kg~m^{-3}$ while the radius of the particle is determined from the linewidth \cite{RahmanSciRep2016} and has an uncertainty of $\approx 4\%$. Fig. \ref{fig4}b are histograms of particle position obtained from the time traces after the voltage to position conversion \cite{Gieseler2012,RahmanRevScIns2018,Neuman2004}. As expected, the $z-$axis has the widest distribution due to the larger spot size along this direction, while the $x-$axis is the narrowest. A measurement of the position distribution, $p(r)$, can be used to reconstruct the potential $U(r)$ assuming a Boltzmann distribution \cite{RicciNatComm2017,RondinNatNano2017}, where $p(r)=Z^{-1}e^{-U(r)/k_BT}$, and $Z$ is the normalization constant determined from the relation $\int{p(r) dr}=1$. Provided that there are a statistically significant number of data points, the potential $U(r)$ can be recovered from the probability distribution of the experimental position data. Figure \ref{fig4}c shows the potential profiles for the three axes determined using this procedure. These match very well with the quadratic functions (solid red lines). For comparison, in Fig. \ref{fig4}d, we show the potential profiles along the three axes obtained using the same particle levitated by the laser. To use the same particle we transfer it from the SLD beam to the laser beam by gradually increasing the laser power whilst keeping the power of the SLD beam constant. After reaching the desired level of power in the laser beam, we reduce the power of the SLD beam gradually to zero. The particle can be transferred from the laser beam to the SLD beam in a similar manner. This procedure is carried out at a pressure of $\approx 5~$mBar using the PBS and the half-waveplate combination shown in Fig. \ref{fig1}. Furthermore, to smoothly transfer the particle from one beam to other, we ensure the two foci are well overlapped. This is achieved by controlling the divergence of the beams and by initially ensuring spectral overlap of two beams. The trap frequency along the $z-$axis under laser levitation was purposely made equal to that under the SLD levitation (Fig. \ref{fig4}a). This means that the potential profiles in both cases are identical as can be observed from Figs. \ref{fig4}c$\&$d. Note that to achieve the same trap frequency along the $z-$axis, SLD levitation requires $20\%$ more power than laser levitation ($125~$mW at the focus). This extra trapping power to achieve the same trap frequency implies a spread in the focus along the $z-$axis. This confirms our simulation results shown in Fig. \ref{fig2c}c, where under the same trapping power the superluminescent diode creates a shallower potential well and hence a lower trap frequency along the propagation direction. The trap frequencies along the $x~\&~y~$axes under the laser levitation (data are not shown) are lower than those under the SLD levitation. This is expected given that a lower laser trapping power was used \cite{Gieseler2012}. In addition, the linewidth, $\Gamma_0$, of the particle along all three axes, under both laser and SLD levitation, are measured to be equal within experimental error, as expected for harmonic potentials.

\section{Parametric feedback cooling}

Figure \ref{fig6} shows the results of parametric feedback cooling \cite{Gieseler2012} of the center-of-mass motion of a levitated nanoparticle under SLD levitation. In this case, light from the SLD without any filtering and amplification was used to trap the particle. Figure \ref{fig6}a shows the power spectral densities along the three major axes. The top graphs are the PSDs before the feedback cooling is applied. The bottom graphs are with feedback on where the motional energy of the particle is significantly lower. Figure \ref{fig6}b shows the derived temperature of the particle along the three axes as a function of the residual gas pressure inside the vacuum chamber. As the area under the PSD is proportional to the center-of-mass temperature \cite{Gieseler2012}, we determine the temperature at any pressure by comparing its area with that taken at a well defined temperature of $295~$K when the particle is trapped without feedback cooling at a pressure of $5~$mBar. Under parametric feedback cooling the energy of the particle along all three axes is reduced approximately by two orders of magnitude from the initial temperature. The lowest temperature of $470~$mK along the $x~$axis is reached at a pressure of $9\times 10^{-6}~$mBar. This is comparable to similar laser based cooling experiments \cite{Gieseler2012,Vovrosh2017}.

\section{Levitation with an optical potential created by the filtered SLD}

To demonstrate the creation of a anharmonic potential we now filter the broadband light from the SLD using the notch filter. Figure \ref{fig5}a shows one such intensity profile where the width of the notch filter was $\approx 10~$nm. Fig. \ref{fig5}b shows the potential profile along the $z-$axis reconstructed from the position histogram as outline above. Here, the black dots are the experimental data points. Asymmetry in the potential profile is immediately visible. In particular, the overall potential is tilted towards the right. At $|z|=1350~$nm, the difference in the height between the two sides of the potential well is $\approx 980~$K. To illustrate this anharmonicity, we fit a quadratic function (solid red line) to the experimental data. As expected, it deviates from the data points significantly. For a good fit, a polynomial consisting terms up to the $9^{th}$ order (solid green line) are essential. Of significance are the odd powers which relate to the asymmetry that can be clearly observed in Fig. \ref{fig5}b. These coefficients are contained in the footnote \footnote{$a_0=2.361,~a_1=0.02788,~ a_2= 0.000972,~ a_3=-1.472\times 10^{-7},~ a_4=2.191\times 10^{-10}, ~ a_5=2.212\times 10^{-13},~a_6=-3.033\times 10^{-16},~ a_7=-6.762\times 10^{-20},~a_8=1.544\times 10^{-22},~ a_9 =-4.143\times 10^{-26}$.}. In the experimental data, the asymmetry in the overall potential arises from the deep and the shallow potentials created by the spectral peaks at $1058~$nm and $1070~$nm (Fig. \ref{fig5}a). In particular, the depth of each potential well is roughly $\propto P/\lambda^2$, where $P$ is the trapping power and $\lambda$ is the wavelength of the trapping light. As a result, due to the higher (lower) power and the shorter (longer) wavelengths, light centered around $1058~$nm ($1070~$nm) forms a tighter (shallower) trap. This means that the potential for $z<0$ is expected to rise faster than for $z>0$. Our simulation (blue dashed line) agrees qualitatively with the experimental data and discrepancies between the two can be attributed to our model which does not explicitly model propagation through an aspheric trapping lens but utilizes a simple lens whose focal length changes according to its dispersion of $150~$nm per nm change in the wavelength. The inset in Fig. \ref{fig5}b shows the separate wells that the two peaks in the intensity profiles (Fig. \ref{fig5}a) would separately form. The potential well labeled with $L$ is associated with the short wavelength peak ($1058~$nm) and is deeper than the well $R$ created by the long wavelength peak centered at $1070~$nm.

\section{Conclusions and Outlook}
We have demonstrated optical levitation using a broadband superluminescent diode and have shown that this source can be used to form a deep optical harmonic potential that can be used for parametric feedback cooling of levitated nanoparticles. By using the combination of an optical amplifier, and a simple tunable notch filter, we can use the inherent chromatic aberration of the lens combined with the broadband nature of the light to create anharmonic potentials. By using an amplifier in saturation, a constant trapping power can be maintained even when the spectral profile of the input to the amplifier is modified or attenuated. Although, not demonstrated here, both double well and quartic potentials appear feasible using lenses with higher chromatic aberrations or by using a spectrally broader SLD source, both of which are currently commercially available. While a single beam can be rapidly scanned in the focal plane to create a time-averaged anharmonic potential \cite{BerutNature2012}, another weaker beam must be used for detection. In our case both can be produced with a single spectrally filtered SLD beam. In addition, as we can use the same high intensity beam for balanced detection we could, in principle, achieve much better signal-to-noise ratios which is a key ingredient for the center-of-mass feedback cooling. For the case where two beams are used to create anharmonic potentials \cite{RondinNatNano2017}, such as a double well potential, yet another beam must be used for detection and this is known to suffer from detection non-linearities since the non-linear potential in the focal plane is extended. In our case the non-linear potential is not in the focal plane but along the propagation direction and we can again use the same beam for the detection of the motion. These are subtle, but significant advantages, when trying to retrieve quantitative measurements of the motion of the particle.  

Dynamic diffractive elements, such as digital mirror devices and spatial light modulators, are ideal for creating similar arbitrary optical potentials. However, these types of modulators are currently limited to modulation speeds of less than $100~$kHz which is often not sufficient for dynamically changing potential landscapes in the underdamped environment of a vacuum levitation experiment. In our method, rapid modulation of the spectral mask on sub-microsecond ($10~$MHz) timescales appears feasible using either an electro-optic or acousto-optic deflector/modulator. This approach will allow modulation of localised parts of the potential landscape instead of the whole well depth, as is done conventionally in feedback cooling. This opens the way for new cooling protocols and for creating modulated potentials that are currently used in thermodynamics experiments \cite{BerutNature2012,RondinNatNano2017}. As ground state cooling in levitated systems has recently been achieved, the creation of non-classical states of motion is now being pursued. This includes the creation of centre-of-mass Schrodinger cat states, which will most probably require a anharmonic potential that can be rapidly switched. We envisage that the fast switching and shaping of a nonlinear potential as a function of time could be used for coherent inflation, where cooling to close to ground state is carried out, followed by a rapid transformation to an inverted potential that is used to speed-up the expansion of the wavefunction. Such a protocol is seen as a promising approach to prepare larger quantum superpositions allow us to reduce the effects of motional decoherence. \cite{Romero_Isart_2017}.  

Lastly, as levitated systems in high vacuum are subject to the photon statistics of the trap light via photon recoil, the SLD and its tunable spectral profile, offers the possibility to study the mechanical effects of incoherent thermal light on the levitated particles \cite{Jain2016,NovotnyPRA2017}.


%

\end{document}